# Cold electron beams from cryo-cooled, alkali antimonide photocathodes

*L. Cultrera\*, S. Karkare, H. Lee, X. Liu, I. Bazarov, B. Dunham*

*Cornell Laboratory for Accelerator-Based Science and Education, Cornell University, Ithaca 14853 NY*

*Abstract:*

In this letter we report on the generation of cold electron beams using a $Cs_3Sb$ photocathode grown by co-deposition of Sb and Cs. By cooling the photocathode to 90 K we demonstrate a significant reduction in the mean transverse energy validating the long standing speculation that the lattice temperature contribution limits the mean transverse energy or thermal emittance near the photoemission threshold, opening new frontiers in generating ultra-bright beams. At 90 K, we achieve a record low thermal emittance of 0.2 $\mu$m (rms) per mm of laser spot diameter from an ultrafast (sub-picosecond) photocathode with quantum efficiency greater than $7 \times 10^{-5}$ using a visible laser wavelength of 690 nm.

The quest for photocathodes that generate electron beams with increased brightness to drive X-Ray Free Electron Lasers (X-FEL) [1], Energy Recovery Linacs (ERL) [2], electron cooling of hadron beams [3], inverse Compton scattering [4] and ultrafast electron diffraction (UED) [5] experiments has recently received much attention from the scientific community, resulting in a stronger interaction between accelerator and solid state physicists trying to identify suitable materials with improved performance for future accelerators and novel applications [6].

From the point of view of the electron source for an accelerator device, the photocathode has to satisfy several, often conflicting, requirements: high quantum efficiency (QE); low thermal emittance or mean transverse energy (MTE); prompt response time; and photocathode longevity. As of today no photocathode material is able to fulfill all these requirements and application specific tradeoffs have to be made to select an acceptable photocathode material.

High QE is important mostly when defining the specifications for the drive laser: higher QE means that less laser power is needed for the same extracted beam current. For high current (~100 mA) applications, like ERL's [2] and electron cooling [3], QE's of a few percent or more in the visible range of the spectrum are needed in order to maintain the average laser power within a few tens of Watts [2]. However, for most other applications such as single pass FEL's or UED setups, where the current requirement is in the 100 µA range [1] and/or the maximum charge per bunch extracted should be kept to a small to avoid charge emittance degradation and QE in the $10^{-5}$ range in visible or UV light suffices. Along with sufficient QE, most accelerator applications require photocathodes to have a sub-picosecond response time and robustness to vacuum conditions in order to operate without significant QE degradation.

For a given bunch charge, the maximum electron beam brightness achievable from the photoemission source depends only on the MTE and the electric field at the cathode surface [7,8,9] and the transverse coherence length, which sets the upper limit to the unit cell size of the crystal that can be imaged using UED setups, depends inversely on the thermal emittance[5]. The thermal emittance is determined by the laser beam size and the MTE of the emitted electrons through the relation:

$$\varepsilon_{n,x} = \sigma_{l,x}\sqrt{\frac{MTE}{m_e c^2}} \qquad (1)$$

where $\varepsilon_{n,x}$ is the normalized transverse emittance in the x plane, $\sigma_{l,x}$ is the rms laser spot size, $m_e$ is the electron mass, and c is the speed of light. Lowering the MTE will increase the beam brightness for FEL other light source applications and increase the transverse coherence lengths for UED applications extending the frontiers of these applications.

Most of the photocathodes used today provide MTEs of a few hundred meV that translate to a transverse coherence length of a couple of nm allowing UED to be performed on samples with unit cells smaller than 1 nm. Reducing MTE below 10 meV transverse coherence lengths larger than 10 nm are achievable allowing UED of larger unit cells like that of protein crystals.

The question of the lowest MTE achievable is an important basic question at the interface of solid state and beam physics. Disorder induced heating of electrons after emission theoretically limits MTEs to 1-2 meV [10]. Smallest measured MTEs are close to 25 meV from GaAs activated using Cs and $NF_3$ [11] and from antimony films [12]. However, GaAs cathodes have a very long response time under infrared illumination and antimony films have extremely low QE (<$10^{-6}$ in the UV range) near the threshold making them impractical for ultrafast accelerator applications.

In this letter, we report on MTE from a Cs$_3$Sb cathode at room temperature and at a cryogenic temperature of 90 K at a near threshold wavelength on 690 nm. At room temperature the cathode delivers a MTE of 40 meV with a QE in the 10$^{-3}$ range and at 90 K the cathode delivers a record low MTE of 22 meV with a QE of 7x10$^{-5}$ under low electric fields of ~100 V/m. Our results not only show that Cs$_3$Sb is an ultrafast, low MTE cathode for FEL and UED applications, but also show that the MTE obtained from these cathodes, when operated near threshold, is indeed limited by the lattice temperature and reducing the lattice temperature can open new frontiers in generating ultra-bright beams for FEL and UED applications.

The experiments have been performed at the photocathode laboratory at Cornell University [13]. The UHV installation has been recently upgraded with a compact high voltage, up to 20 kV, photoelectron gun which includes a solenoid and 2 pairs of corrector coils (figure 1). We'll refer to this electron gun with the name of Transverse Energy meter (TEmeter) from now on.

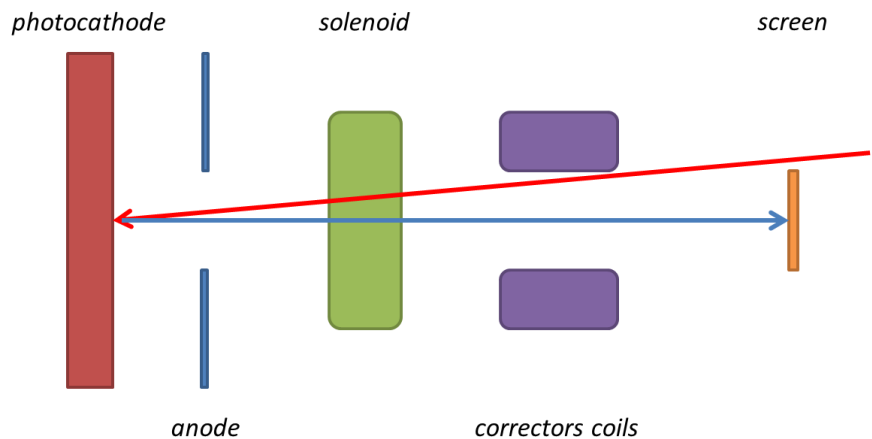

Figure 1. TEmeter scheme: laser (red) and electron beam (blue) paths are indicated.

The photocathode holder is connected to a liquid nitrogen reservoir, allowing the cooling of the photocathode from room temperature down to ~90 K. The 50 mm stainless steel circular anode has a 12 mm diameter hole allowing light generated using laser diodes or by an optical system comprising a lamp and a monochromator be sent to the photocathode surfaces through a UHV window with ~6° angle with respect to the axis of the electron gun. Electrons are accelerated by the electric field generated between the negatively biased photocathode surface and the grounded anode, and are imaged onto a Ce-doped YAG scintillating screen. A CCD camera is used to measure the beam size. The YAG screen is coated with 7.5 nm of titanium to prevent surface charge accumulation.

A $Cs_3Sb$ cathode has been grown on a p-doped Si(100) substrate while cooling from 130 °C to 50 °C, by co-evaporating Sb and Cs with respective fluxes of $3\times10^{11}$ and $1\times10^{12}$ atoms cm$^{-2}$ s$^{-1}$ (see figure 2a) until the QE reached a plateau at 0.05 at 532 nm (see figure 2b).

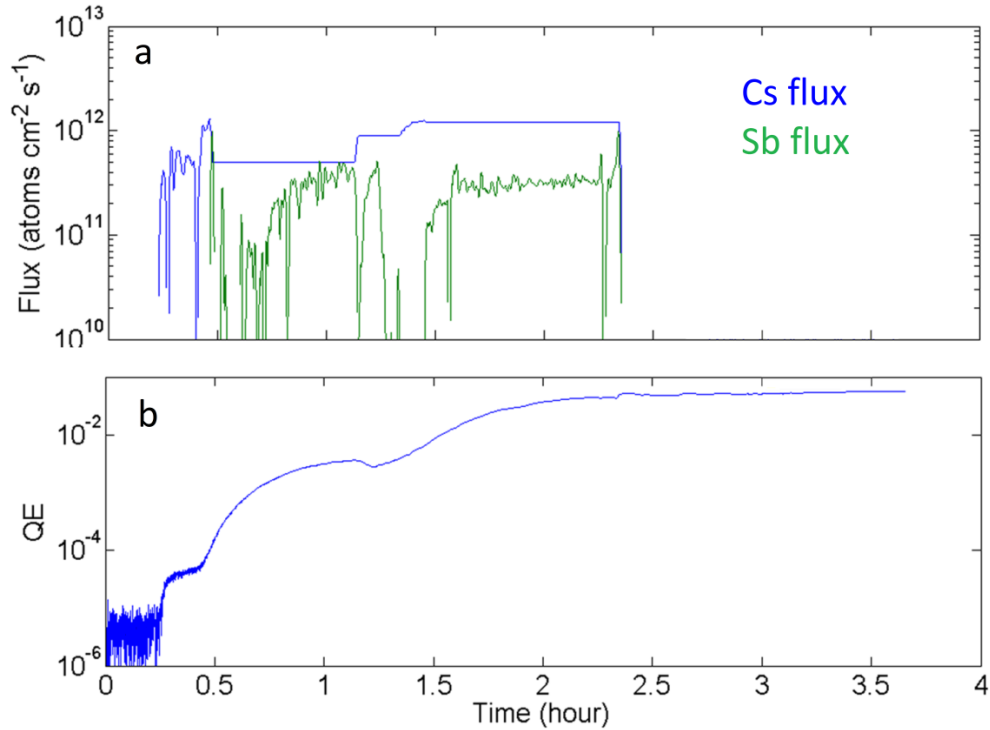

Figure 2. (a) Fluxes of Cs and Sb vapors and (b) QE at 532 nm during the growth of the Cs3Sb photocathode.

The cathode was then moved from the growth chamber to the high voltage (350 kV) electron gun of the Cornell-ERL injector prototype [14] under UHV by using a vacuum suitcase without exposing it to air.

QE and thermal emittance at 690 nm were measured at room temperature using a laser diode with the solenoid scan technique [15]. Results are reported in figure 3 as function of the gun voltage: while QE is seen to increase from $7\times10^{-4}$ to $1.5\times10^{-3}$ due to the Schottky lowering of the work function, the beam emittances and hence MTEs does not show any remarkable field dependence.

QE dependence on the applied electric field can be expressed as

$$QE = a\left(hv - \varphi + b\sqrt{\beta E}\right)^2 \tag{2}$$

Where $a$ is a material dependent constant, $hv$ is the photon energy, $\varphi$ is the threshold wavelength for the photoemission, $b$ is defined as $b = \sqrt{e/4\pi\varepsilon_0}$, $E$ is the applied electric field and $\beta$ its enhancement factor. $\beta$ can be deduced from a linear fit of $QE^{0.5}$ as function of $E^{0.5}$ resulting to be 3.7.

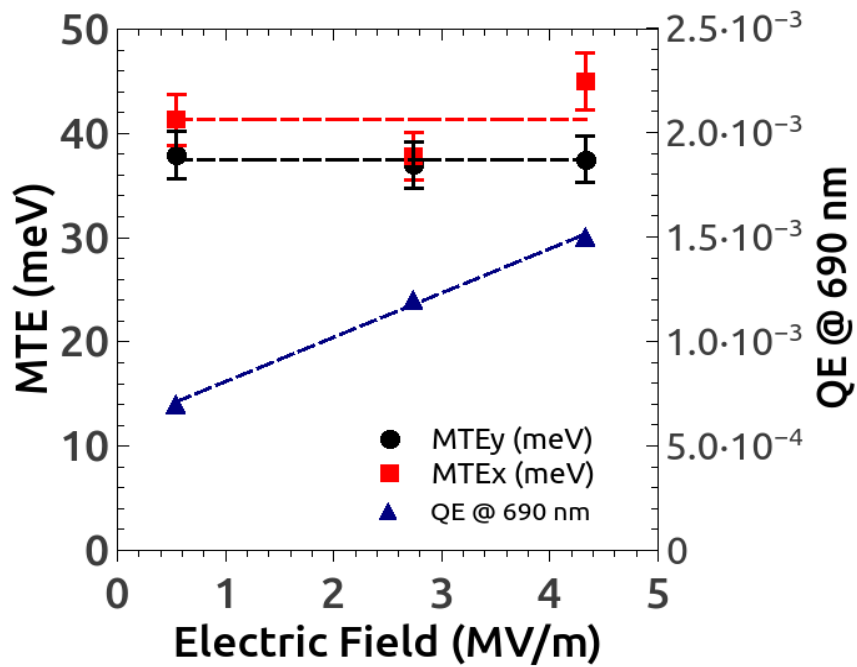

Figure 3. Cs$_3$Sb photocathode QE and thermal emittance at 690 nm as measured in the injector DC gun.

The cathode was then moved back to the photocathode laboratory using the vacuum suitcase and once installed into the TEmeter, it was subjected to few cycles of cooling to cryogenic temperatures (90 K) and back to room temperature (300 K). Thermal emittance was estimated using two different methods: the solenoid scan and the free expansion of the electron beam. Data analysis is performed by means of linear optics transfer matrices as illustrated in ref. 15 and for the free expansion method the solenoid field is set to zero.

Thermal emittances were determined using the solenoid scan technique with an rms laser spot size of 60±3 and 64±3 µm respectively for x and y direction, using photocurrent intensities in the range of 1 to 2 nA to avoid space charge and for three different gun voltages (5, 7 and 9 kV) corresponding the electric field gradients of 0.88, 1.23 and 1.58 MV/m at 300 K and 90 K as reported in figure 4.

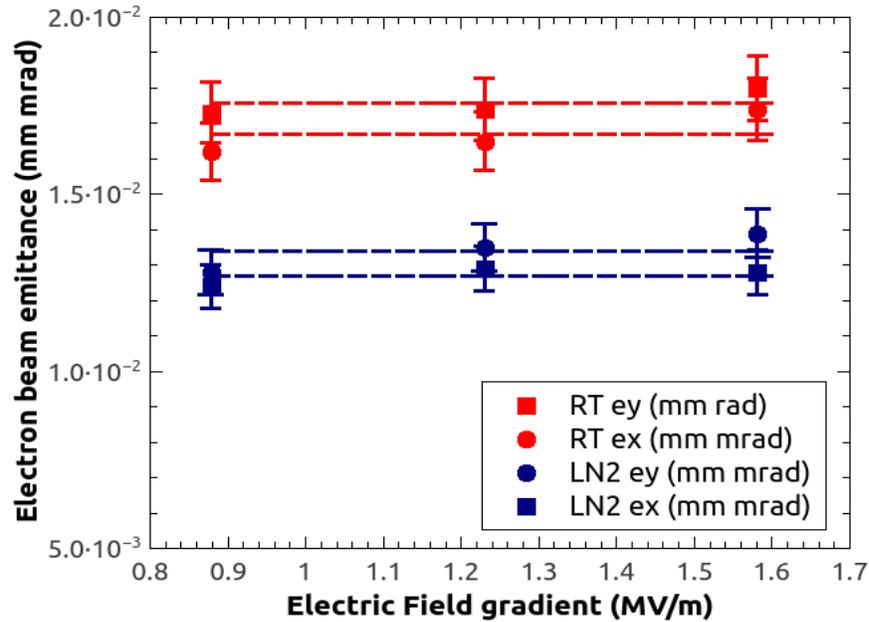

Figure 4. Solenoid scan measured normalized emittance in the TEmeter at 300 K (red) and at 90 K (blue).

From a linear fit it is deduced that the MTEs of the electron beam at 300 K and at 90 K are of 40±2 and 22±1 meV respectively. The agreement between measurements performed at room temperature in the TEmeter and in the ERL injector prototype DC gun is noteworthy.

Solenoid scan measurements with laser spot size larger than 60 um rms have been found to be affected from solenoid aberration most likely due to a relatively large electron beam size inside the solenoid magnetic field. For these reasons we performed additional measurements by leaving the electron beam going under a free expansion using different laser beam sizes and different electric field intensities at the cathode surface varying from 0.5 to 3.4 MV/m. The results of the measurements are summarized in figure 5 where the emittance of the beam is reported as function of the initial laser spot size at the cathode surface.

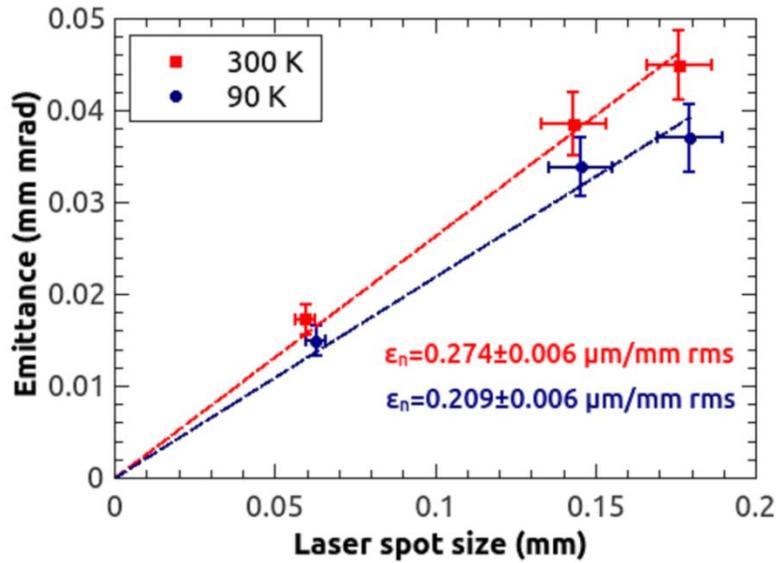

**Figure 5.** Thermal emittance is deduced from the linear fit of the beam emittances as function of laser spot size.

From the linear fit of the beam normalized emittance as function of the laser beam spot size reported in figure 5 the thermal emittance of 0.274±0.006 µm/mm rms (38±2 meV) and 0.209±0.006 µm/mm rms (22±1 meV) are estimated at 300 K and 90 K respectively.

The Spectral response of the $Cs_3Sb$ cathode was measured biasing the cathode with -18 V at both temperatures and reported in figure 6. The influence of Schottky effect for this measurement is negligible.

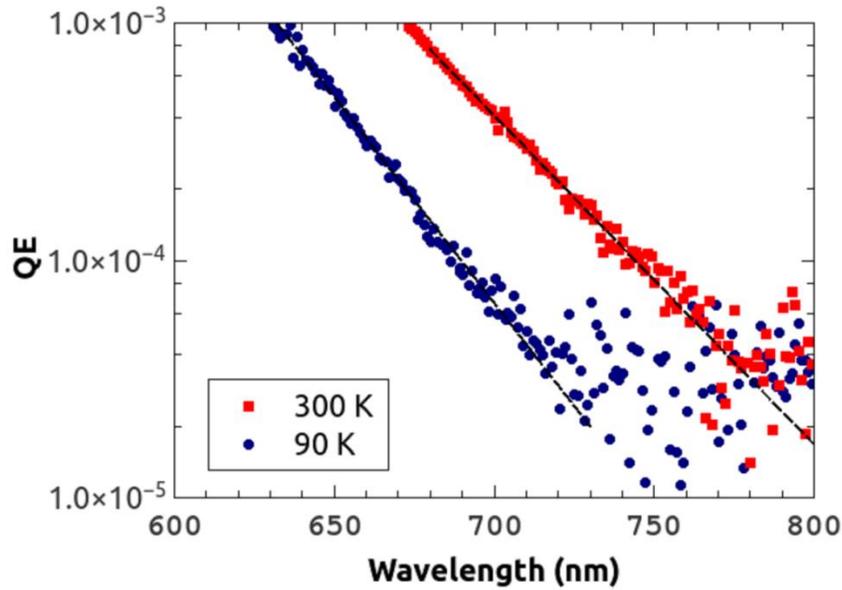

**Figure 6. Spectral response at 300 K and 90 K.**

The extrapolation of the spectral response data allows estimating the threshold $\varphi$ for photoemission which, within our experimental sensitivity, was ~1.59 eV and ~1.72 eV at 300 K and at 90 K respectively.

In a very simple model assuming isotropic photoemission with electron energy uniformly distributed in the interval 0 and $h\nu - \varphi$ it can be shown that MTE simply reduces to $(h\nu - \varphi)/3$ [15]. This simple evaluation does not include the field dependent contribution of the Schottky work function lowering that, using the estimated field enhancement coefficient obtained from QE measurement, can be as large as ~135 meV for the largest value (3.4 MV/m at 20 kV) of the electric field gradient in the TEmeter. The predictions of such simple model yield expected values for MTEs of ~70 meV and ~25 meV for 300 K and 90 K respectively. If the contribution of the Schottky effect is included, the electric field dependence should be observed on MTEs and values as large as ~115 meV should be observed at 300K and at 3.4 MV/m. In addition to that, it is well known that the electron beam thermal emittance is affected by the cathode surface roughness. Electrons trajectories can be affected by the photocathode surface roughness through two main processes: one is due to the relative geometric orientation of local surfaces from where the electron originates with respect to the nominal photocathode normal, the other mechanism is due to a transverse kick imparted to photoelectrons near the photocathode surface by local transverse electric field components, in this latter case it can be shown that MTE shows a linear dependence on the electric field [16]. This whole picture looks in contrast with our smaller measured MTEs indicating that a model simply scaled on electrons excess energy cannot explain our experimental result.

Already in 1958, Spicer observed photoemission from alkali antimonide photocathodes with photons having energy lower than the sum of the band gap (Eg) and electron affinity (Ea)[17]. This observation was interpreted as photoemission from filled donor levels lying near the Fermi level within the gap of the semiconductor. The number of electrons filling these states and contributing to the photoemission process at photon energies lower than the Eg+Ea (~2 eV) level is strongly related to the temperature of the sample trough the Fermi-Dirac distribution. Spicer also observed that, as expected from its interpretation, at cryogenic temperatures the QE of the alkali antimonides photocathodes strongly decreased because of the reduction of the number of electrons filling these donor levels having energies large enough to be excited above the photoemission threshold [17]. Our observation of QE increasing and decreasing as temperature changes is consistent with those results.

A recent model, which includes the effect of the finite temperature on the Fermi Dirac distribution of the electrons in metals, shows that the thermal emittance due to photoelectrons can be expressed as [18]:

$$\epsilon_{n,x} = \sigma_{l,x} \sqrt{\frac{kT}{m_e c^2}} \sqrt{\frac{Li_3\{-exp[\frac{e}{kT}(h\nu-\varphi)]\}}{Li_2\{-exp[\frac{e}{kT}(h\nu-\varphi)]\}}} \qquad (3)$$

Where $Li_n$ is a polylogarithm function defined as [18]:

$$Li_n(z) = \frac{(-1)^{n-1}}{(n-2)!} \int_0^1 \frac{1}{t} log(t)^{n-2} log(1-zt) dt \qquad (4)$$

The presence of filled donor states within the energy gap allows us to use the same formula to estimate the emittance and hence the MTE of electron beam generated during our measurements. In our experimental conditions the photon energy (~1.8 eV) is smaller than the sum of energy gap and electron affinity of the material (~2 eV) and under this circumstances the ratio of the two poly-logarithm functions in expression (2) tends to 1, thus the measured MTE in the absence of other effects should equal $kT$ yielding MTEs of 25 meV and 8 meV for cathode temperature of 300 K and 90 K respectively. This result does not depend on the electron affinity as long as the photon energy is smaller than the effective work function, which includes the lowering induced by the Schottky effect. The absence of a measurable increase of the thermal emittance with electric field in our result supports this assumption. Nonetheless, the values of MTEs measured at 300 K and 90 K are larger than the expected values. The increase of the MTE values can be attributed to the surface roughness. Indeed, considering a simplified two dimensional roughness model as [16]

$$z(x) = a \cos\left(2\pi \frac{x}{\lambda}\right) \quad (5)$$

Where $a$ is the amplitude and $\lambda$ the period of the modulation of the surface roughness the thermal emittance increase $\varepsilon_{th,2D}$ can be expressed as

$$\varepsilon_{th,2D} \leq \varepsilon_{th}\sqrt{1 + 6\left(\frac{\pi a}{\lambda}\right)^2} \quad (6)$$

The surface of alkali antimonide cathodes has been reported to have a roughness on the order of 25 nm rms with period of 100 nm [19]. For such surface roughness parameter the term under the square root equals 2.16. This estimated contribution alone exceeds and can be sufficient to explain the differences from the expected and measured values for the MTEs at both 300 K and 90 K. This contribution depending only on the geometry of the surface and not on the electric field well fit with our experimental observations.

Exploring the photoemission properties of a bi-alkali antimonide photocathode near the emission threshold with visible light at 690 nm and at cryogenic temperatures we have demonstrated that sub-room-temperature electrons (~0.2 µm/mm rms) can be produced with QE (at least 7x10$^{-5}$) comparable to that of commonly used metals, which, however, require UV excitation light. Electron beams brightness can be improved usingalkali antimonides as compared to the metal photocathodes having similar QE but necessitating ultraviolet photons to overcome a larger work-function. The copper cathode operating in the LCLS photoinjector has a typical QE of $1\times10^{-4}$ and a thermal emittance of ~0.9 μm/mm rms when illuminated with UV light at 253 nm [20]. MTE's comparable to the ones we reported here can be achieved using Magneto Optical Trapping and excitation of Rb atoms as described in [21], but due to practical limitation in achieving higher density of Rb atoms within the interaction region with the laser beams, the charge per bunch so far has been limited to few tens of fC.

In summary, we have reported on the generation photoelectrons beams with sub-room-temperature MTE from $Cs_3Sb$ photocathodes suitable for applications requiring ultra-bright and ultra-short bunches. Even without the cryogenic cooling, the beam brightness will be better than previously demonstrated photocathodes. We note that the surface roughness is a noticeable contributor to the transverse momentum spread of the electrons for this type of photocathodes. Therefore, work is underway to realize atomically flat cathodes of this family. The cooling of the photocathode to even lower temperatures than in our present setup may eventually lead to the thermal emittances and MTE of a few meV especially if the cathode surface roughness contribution to emittance growth can be minimized.

This work has been funded by the NSF under Grants no. PHY-1416318, DMR-0807731 and by the DOE under Grant no. DE-SC0003965.

**References**
1. W. Ackermann et al., *Nat. Photonics* **1**, 336 (2007)
2. S. M. Gruner, D. Bilderback, I. Bazarov, K. Finkelstein, G. Krafft, L. Merminga, H. Padamsee, Q. Shen, C. Sinclair and M. Tigner, *Rev. Sci. Instrum.* **73**, 1402 (2002)
3. I. Ben-Zvi et al., *Nucl. Instrum. Methods Phys. Res., Sect. A* **532**, 177 (2004)
4. S. Boucher, P. Frigola, A. Murokh, M. Ruelas, I. Jovanovic, J. B. Rosenzweig, and G. Travish, *Nucl. Instrum. Methods Phys. Res., Sect. A* **608**, S54 (2009)
5. T. Van Oudheusden, E. F. de Jong, S. B. van der Geer, W. P. E. M. Op ' t Root, O. J. Luiten, and B. J. Siwick, *J. Appl. Phys.* **102**, 093501 (2007).

6. D. H. Dowell, I. Bazarov, B. Dunham, K. Harkay, C. Hernandez-Garcia, R. Legge, H. Padmore, T. Rao, J. Smedley, and W. Wan, *Nucl. Instrum. Methods Phys. Res., Sect. A* **622**, 685 (2010)
7. I.V. Bazarov, B.M. Dunham, C.K. Sinclair, *Phys. Rev. Lett.* **102** 104801 (2009)
8. C. Gulliford, A. Bartnik, I. Bazarov, L. Cultrera, J. Dobbins, B. Dunham, F. Gonzalez, S. Karkare, H. Lee, H. Li, Y. Li, X. Liu, J. Maxson, C. Nguyen, K. Smolenski, Z. Zhao, *Phys. Rev. Spec. Top.-Accel. Beams* **16,** 073401(2013)
9. C. Gulliford, A. Bartnik, I. Bazarov, B. Dunham, L. Cultrera, *Appl. Phys. Lett.***,** submitted
10. J. M. Maxson, I. V. Bazarov, W. Wan, H. A. Padmore, and C. E. Coleman-Smith, New J. Phys. 15, 103024 (2013).
11. I.V. Bazarov, B.M. Dunham, Y. Li, X. Liu, D.G. Ouzounov, C.K. Sinclair, F. Hannon, T. Miyajima, *J. Appl. Phys.*, **103** (2008) 054901
12. J. Feng, J. Nasiatka, W. Wan, T. Vecchione, and H. A. Padmore, *Rev. Sci. Instrum.,* 86, 015103 (2015).
13. L. Cultrera et al., *Proceedings of IPAC2012*, New Orleans, Louisiana, 2012, p. 2137 (2013)
14. B. Dunham, A. Bartnik, I. Bazarov, L. Cultrera, J. Dobbins, G. Hoffstaetter, B. Johnson, R. Kaplan, V. Kostroun, S. Karkare, Y. Li, M. Liepe, X. Liu, F. Loehl, J. Maxson, P. Quigley, J. Reilly, D. Rice, D. Sobol, E. Smith, K. Smolenski, M. Tigner, V. Veshcherevich, Z. Zhao, *Appl. Phys. Lett.* **102** 034105(2013)
15. I. V. Bazarov, L. Cultrera, A. Bartnik, B. Dunham, S. Karkare, Y. Li, X. Xianghong, J. Maxson, and W. Roussel, *Appl. Phys. Lett.* **98**, 224101 (2011)
16. H. J. Qian, C. Li , Y. C . D u, L. X. Yan, J. F. Hua, W. H. Huang, C. X . Tang, *Phys. Rev. Spec. Top.-Accel. Beams* **15** , 040102 (2012)
17. W.E. Spicer, *Phys. Rev.* **112**, 114 (1958)
18. T. Vecchione, D. Dowell, W. Wan, J. Feng, H. A. Padmore, *Proceedings of FEL 2013*, New York, NY, p. 424 (2013)
19. S. Schubert, M. Ruiz-Osés, I. Ben-Zvi, T. Kamps, X. Liang, E. Muller, K. Müller, H. Padmore, T. Rao, X. Tong, T. Vecchione and J. Smedley, APL Materials **1**, 032119 (2013)
20. F. Zhou, A. Brachmann, F-J. Decker, P. Emma, S. Gilevich, R. Iverson, P. Stefan, and J. Turner, *Phys. Rev. Spec. Top.-Accel. Beams* **15**, 090703 (2012)
21. W. J. Engelen, M. A. van der Heijden, D. J. Bakker, E. J. D. Vredenbregt and O. J. Luiten, Nature Communications **4**, 1693 (2013)